\documentclass[prl,aps,amsfonts,amssymb,draft,floats,
twocolumn,showpacs]{revtex4}
\usepackage{epsf}

\newcommand{\be}{\begin{equation}}
\newcommand{\ee}{\end{equation}}
\newcommand{\one}{\openone}
\newcommand{\vS}{\mathbf{S}}
\newcommand{\vH}{\mathbf{H}}

\renewcommand{\vr}{\mathbf{r}}
\newcommand{\spacing}{\delta}

\begin{document}

\title{Fluctuations of $g$-factors in metal nanoparticles: Effects of 
electron-electron
interaction
and spin-orbit scattering}

\author{Denis A. Gorokhov\footnote{e-mail: gorokhov@ccmr.cornell.edu} 
and Piet W. Brouwer}

\affiliation{Laboratory of Atomic and Solid State Physics,
Cornell University, Ithaca, NY 14853-2501, USA}

\begin{abstract}
We investigate the combined effect of spin-orbit
scattering and electron-electron interactions on the
probability distribution of $g$-factors of metal 
nanoparticles. 
Using random matrix theory, we find that even a relatively small 
interaction strength
significantly
increases $g$-factor fluctuations
for
not-too-strong spin-orbit scattering (ratio of spin-orbit
rate and single-electron level spacing $1/\tau_{\rm so} \spacing
\lesssim 1$), and leads to the possibility to observe $g$-factors larger 
than two.
\end{abstract}


\maketitle



Electronic properties of metal nanoparticles can be studied on the
single-electron level using ``tunneling spectroscopy'', the
measurement of the conductance of a metal particle
coupled to source and drain electrodes via tunneling contacts
\cite{kn:vondelft2001}. These measurements have revealed important
insights into the nature of the electronic ground and 
of individual excited
states of normal metal, ferromagnetic, and superconducting
nanoparticles
\cite{kn:black1996,kn:gueron1999,kn:davidovic1999,kn:petta2001,kn:petta2002}.
Although tunneling spectroscopy involves processes in which only
a single electron is added to or removed from the metal particle,
the location of conductance peaks gives information about
many-electron energy levels and, hence, about the role of
electron-electron interactions. 


The combination of electron-electron interactions and 
mesoscopic fluctuations of the density of states of a normal-metal
nanoparticle can lead to a many-electron ground state 
with nontrivial spin $S=1$, $S=3/2$, or even larger 
\cite{kn:brouwer1999b,kn:baranger2000}. (Without interactions,
the ground state spin is always $S=0$ or $S=1/2$.)
In principle, the spin of a many-electron state $|k\rangle$ can be
measured via the derivative of the energy
$E_k$ versus an applied magnetic field $H$, which is parameterized 
using the ``$g$-factor'' $g_k$,
\be
  \left. \frac{\partial E_k}{\partial H} \right|_{H \to 0} 
  = \pm \frac{1}{2} g_k \mu_{B},\ \
  \mu_B = \frac{|e| \hbar}{2 m c}.
  \label{eq:gfactors}
\ee
If spin is a good quantum number, the positions of tunneling 
spectroscopy conductance peaks are unaffected by a nontrivial value
of $S$,
since tunneling spectroscopy measures
differences of $g$-factors of many-electron states for
which the electron number differs by one. Since the spin of a nanoparticle
changes by $1/2$ upon addition or removal of an electron, all observed
(differences of)
$g$-factors are equal to two, irrespective of the individual 
$g$-factors of the two many-electron levels participating in the
transition. 

In this letter, we consider $g$-factor differences
measured in tunneling spectroscopy
in the presence of both electron-electron interactions and
spin orbit scattering. 
When spin and orbital degrees of freedom are coupled, randomness
in the orbital part of wavefunctions is passed on to the spin part. 
Randomizing the electron spin lifts the 
``selection rule'' that prohibited the observation of $g$-factors
larger than two in the absence of spin-orbit scattering. It also 
leads to a decrease of $g$-factors, a suppression of the long-range
exchange interaction (which is responsible for the high-spin states), 
and level-to-level fluctuations
of $g$-factors \cite{kn:brouwer2000,kn:matveev2000}. 
As we show here, selection rules are lifted already for a small
spin-orbit scattering rate $1/\tau_{\rm so} \lesssim \spacing$,
$\spacing$ being the mean spacing between single-electron
levels,
whereas the decrease of $g$-factors and the suppression of the exchange
interaction become effective
at a larger spin-orbit rate $1/\tau_{\rm so} \gtrsim \spacing$ only,
leaving a substantial parameter window where 
$g$-factor differences larger than two can be observed. Such 
large $g$-factors are
a true many-electron phenomenon since, without interactions, all
measured $g$-factors correspond to single-electron levels and
are always $\le 2$ \cite{foot1}.

{\em Model.}
Without interactions, the single-electron wavefunctions and
energy levels of a metal nanoparticle are described by random
matrix theory. With spin orbit scattering, the appropriate random 
matrix ensemble
interpolates between the Gaussian Orthogonal Ensemble (GOE) and the
Gaussian Symplectic Ensemble (GSE) 
\cite{kn:brouwer2000},
\begin{equation}
  H_0 = H_{\rm GOE} +  H_{\rm so}.
  \label{eq:H0}
\end{equation}
Writing the spin degrees of freedom explicitly, one has
$$
  H_{\rm GOE} = S \otimes \one_2,\ \ 
  H_{\rm so} = \frac{\lambda}{\sqrt{4 N}}
  \sum_{j=1}^{3} i A_j \otimes \sigma_j,
$$
where $\one_2$ is the $2 \times 2$ unit matrix in spin space,
$\sigma_j$ is the Pauli matrix ($j=1,2,3$), $S$ is an 
$N \times N$ real symmetric matrix, $A_j$ is an $N\times N$ real
antisymmetric matrix ($j=1,2,3$), and $\lambda^2 = \pi/\tau_{\rm so}
\spacing$ is the dimensionless spin-orbit scattering rate. 
The elements of the matrices
$S$, $A_1$, $A_2$, and $A_3$ are drawn from independent
Gaussian distributions with zero mean and with equal variances
for the off-diagonal elements. The diagonal elements of $S$
have double variance, whereas the diagonal elements of 
$A_1$, $A_2$, and $A_3$ are zero because of the antisymmetry
constraint. The limit $N \to \infty$
is taken at the end of the calculation. Each eigenvalue 
$\varepsilon_{\mu}$ of $H_0$ is doubly degenerate, with
wavefunctions $\psi_{\mu 1}$ and $\psi_{\mu 2}$ related by
time-reversal.

In normal-metal nanoparticles,
the main contribution to electron-electron interactions is described 
by the ``constant exchange interaction model'' \cite{kn:kurland2000,foot0},
\begin{equation}
  H_{\rm ex} = - J \vS^2, \label{eq:Hex}
\end{equation}
where $\vS$ is the total spin of the particle. The ratio of the
exchange constant $J$ and the mean spacing between single-electron
levels $\spacing$ corresponds to one of the Fermi Liquid constants 
of the metal. For most normal metals one has $0.2 \lesssim J/\spacing 
\lesssim 0.4$, in agreement with electron-liquid theory
\cite{kn:hedin1969}, although smaller and larger values occur as
well (see the discussion at the end of this letter).
Combining the constant exchange interaction (\ref{eq:Hex}) and the
single-electron Hamiltonian (\ref{eq:H0}), and including
the Zeeman coupling to a magnetic field $\vH$ in the $z$ direction
\cite{foot1}, one has
\be
  {\hat H} = \sum_{\mu} \varepsilon_{\mu} 
  (\hat \psi^{\dagger}_{\mu1} 
  \hat \psi^{\vphantom{\dagger}}_{\mu1} +
  \hat \psi^{\dagger}_{\mu2} 
  \hat \psi^{\vphantom{\dagger}}_{\mu2})
  - J {\bf {\hat S}}^2 - 2 \mu_B H S_z,
\label{hamiltonian}
\label{half_of_Hamiltonian}
\ee
where the first term comes form the diagonalization
of the single-electron Hamiltonian (\ref{eq:H0}). 

If the number of electrons $N_e$ is even, all many-electron states are
non-degenerate in the presence of spin-orbit scattering.  Hence all
even-electron states have $g$-factors equal to zero.  With spin-orbit
scattering, odd-electron states are twofold degenerate (Kramers'
degeneracy). Since tunneling spectroscopy measures differences of 
$g$-factors for many-electron states with $N_e$ and $N_e + 1$ electrons, a
$g$-factor measured using tunneling spectroscopy is the $g$-factor of
an odd-electron state if spin-orbit scattering is present.  
We assume that the nanoparticle
relaxes to the even-electron ground state between tunneling events so
that one measures $g$-factors of the odd-electron ground state and
the odd-electron excited states upon increasing the bias voltage. 
In the (generic) case that the two tunneling
contacts between the nanoparticle and the source and drain reservoirs
have very different conductances, the height of a
conductance peak is proportional to the matrix element
\cite{kn:vondelft2001}
\begin{eqnarray}
  w_{k} &=& \sum_{\sigma=\uparrow,\downarrow}
   | \langle N_e+1, k | \hat \psi^{\dagger}_{\sigma}(\vr) 
  | N_e, 0 \rangle |^2 ,
  \label{weights}
  \label{eq:weights}
\end{eqnarray}
where the creation operator $\hat \psi^{\dagger}_{\sigma}(\vr)$
creates an electron with spin $\sigma$ in the grain at the position 
$\vr$ of the point contact with the smaller conductance,
$|N_e, 0 \rangle$ is the even-electron ground state, and
$|N_e+1, k \rangle$ is an odd-electron excited state. Without
spin-orbit scattering, the weights $w_k$ are zero for those
states $|N_e+1, k \rangle$ for which the spin $S_z$ differs by 
more than $1/2$ from the spin of the even-electron ground state
$|N_e, 0 \rangle$. (This is the ``selection rule'' referred to
in the introduction.)

{\it Limit of weak spin-orbit scattering.}  We first address the limit
$\lambda \ll 1$, for which perturbation theory in $\lambda$ is 
possible. In view of the above comments, we need
to consider $g$-factors of odd-electron states only. For an
odd-electron state $|k\rangle$ with spin $S=1/2$ without spin-orbit
scattering, spin-orbit scattering only affects the spin contribution
to the $g$-factor to quadratic order in $\lambda$
\cite{kn:sone1977}. However, all nanoparticles have low-lying 
odd-electron states with spin $S=3/2$ if the exchange 
interaction is present. (There even is a
small but nonzero probability that the odd-electron ground state has
spin $S=3/2$ \cite{kn:brouwer1999b,kn:baranger2000}.)  Spin-orbit
scattering lifts the fourfold degeneracy of an $S=3/2$ odd-electron
state and splits this state into two doublets. As we show below,
spin-orbit scattering determines the $g$-factors of these doublets 
already to {\it zeroth} order in $\lambda$, whereas the matrix elements
determining the corresponding tunneling spectroscopy peak height $w_k$
are nonzero with finite probability.

Labeling the four members
of the $S=3/2$ quadruplet by the $z$ component of the spin,
$S_z = p - 5/2$, $p=1,2,3,4$, the matrix elements of $H_{so}$ can be
arranged in a $4 \times 4$ matrix $V_k$ of the form
\be
V_k = 
\left (
\begin{array}{cccc}
  - a-d   & b        & c          & 0  \\ 
  b^{*}   & - a + d  & 0          & c  \\
  c^{*}   & 0        &  - a + d   & -b \\
  0       & c^{*}    &  - b^{*}   & - a - d 
\end{array}
\right ),
\label{general_form}
\ee
with $a$ and $d$ real numbers and $b$ and $c$ complex numbers.  The
specific form of (\ref{general_form}) follows from time-reversal
symmetry and guarantees that that the eigenvalues of $V_k$ are doubly
degenerate, in accordance with Kramers' theorem.  One has $V_k = 0$ to
first order in $H_{\rm so}$, since spin-orbit scattering does not mix
states with opposite spin belonging to the same energy
level. Calculating $V_k$ to second order in $H_{\rm so}$, we consider
the special case when the $S=3/2$ quadruplet is split by virtual
transitions to one nearby odd-electron state $|l\rangle$ only. We
refer to Ref.\ \onlinecite{kn:gorokhov2003} for a general discussion;
neglecting other states is justified if the energy difference
of the virtual
transition we consider is much smaller than other energy separations. 

Since we are interested in $g$-factors only, 
it is sufficient to calculate the ratios $b/d$ and $c/d$, for which we
find
\begin{eqnarray}
  \frac{b}{d} &=& \frac{(A_1 - i A_2)_{\mu \nu}(A_3)_{\mu\nu} 
  \sqrt{3}}{(A_1)_{\mu\nu}^2 + (A_2)_{\mu\nu}^2 -
  2 (A_3)_{\mu\nu}^2},
  \nonumber \\
  \frac{c}{d} &=& \frac{(A_1 - i A_2)_{\mu\nu}^2 \sqrt{3}}
  {(A_1)_{\mu\nu}^2 + (A_2)_{\mu\nu}^2 -
  2 (A_3)_{\mu\nu}^2},
\end{eqnarray}
where $\mu$ and $\nu$ refer to the two single-electron
levels involved in the virtual transition. The $S=3/2$ quadruplet 
splits into two 
doublets with $g$-factors $g_{k1}$ and $g_{k2}$, with
\begin{eqnarray}
  g_{k1}^2 &=& 
  4+
  \frac{12 ((A_1)_{\mu\nu}^2 + (A_2)_{\mu\nu}^2)}
  {(A_1)_{\mu\nu}^2 + (A_2)_{\mu\nu}^2 + (A_3)_{\mu\nu}^2}, 
  \nonumber \\
  g_{k2}^2 &=& 48 - 3 g_{k1}^2,
\label{simplest_case}
\end{eqnarray}
Whether $g_{k1}$ or $g_{k2}$ corresponds to the
lower lying doublet depends on the relative position and spin of 
the virtual state; 
if the unperturbed $S=3/2$ state is the ground
state, the doublet
with lower energy has $g$-factor $g_{k1}$ \cite{kn:gorokhov2003}.
Using the distribution of the matrices $A_1$, $A_2$, and $A_3$, one 
finds that $g_{k1}$ has the distribution
\begin{equation}
  P_1(g_{k1}) = \frac{1}{2}
  \frac{g_{k1}}{\sqrt{48 - 3 g_{k1}^2}}, \ \
  2 \le g_{k1} \le 4. \label{eq:Pg}
\end{equation}
The second $g$-factor $g_{k2}$ takes values in the interval $0 \le
g_{k2} \le 6$ and is related to $g_{k1}$ via Eq.\
(\ref{simplest_case}). 

To zeroth order in $\lambda$, the tunneling spectroscopy peak heights
(\ref{eq:weights}) are nonzero if the even-electron ground
state $|N_e,0\rangle$ has spin $S=1$. This occurs
with significant probability for $J \gtrsim 0.3 \delta$
\cite{kn:brouwer1999b,kn:baranger2000}, so that
the anomalous $g$-factors $g_{k1}$ and $g_{k2}$ are visible in a
tunneling spectroscopy experiment with finite probability.
Increasing the spin-orbit scattering rate $\lambda$ further
increases the visibility of peaks with the largest $g$-factors by
mixing even-electron states with $S=0$ and $S=1$. 
Indeed, for $J=0.3\delta$ the
average energy difference between the lowest lying $S=0$ and $S=1$
even-electron states is $\sim \delta - 2J = 0.4\delta$, so that
even moderate spin-orbit scattering ($\lambda \gtrsim 0.5
\spacing$) has a matrix element $ \lambda \delta$ between the two 
states that is comparable with the energy difference. 
Inclusion of virtual excitations to other excited 
state changes the formulas for $g_{k1}$ and $g_{k2}$, but
not the conclusion that spin-orbit scattering
affects the $g$-factors of the $S=3/2$ states to zeroth order
in $\lambda$.
\begin{figure}
\epsfxsize= 0.75\hsize
\epsffile{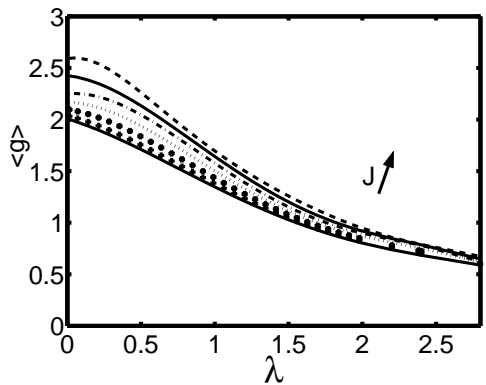}
\caption{Ensemble averaged $g$-factors for $J/\spacing=0$, $0.1$,
$0.2$, $0.3$, $0.4$, $0.5$, and $0.6$.
} 
\label{weighted_average_g_factors}
\end{figure}

{\em Arbitrary spin-orbit scattering rate.} In order to
address the effects of a finite spin-orbit scattering rate, we
have numerically diagonalized the Hamiltonian (\ref{hamiltonian})
for $0 < \lambda < 2.8$. We first diagonalized 
$H_0$ and considered the interaction $H_{\rm ex}$
in the basis of the 92 (76) lowest lying 
many-electron eigenstates of $H_0$ for $N_e$ odd (even). 
We then diagonalized
the remaining many-electron Hamiltonian and calculated the 
$g$-factors $g_k$ of the $M=8$ lowest-lying odd-electron states,
together with peak height $w_k$ for transition from the 
even-electron ground state, see Eq.\ (\ref{eq:weights}). 
The random matrices in our simulation are taken of size $2N=400$
for $2 < \lambda < 2.8$ and of size $2N = 200$
for $\lambda < 2$. We averaged over $300$
realizations of the random matrices, corresponding to a
mesoscopic average over an ensemble
of nanoparticles with equal size and spin-orbit
scattering rate but different disorder configurations.
In the analysis of the numerical data, we discarded all
levels for which the peak height $w_k$ is below a
threshold $w_{\rm tr}$, which we arbitrarily set at 
\begin{equation}
  w_{\rm tr} = 0.1 \times \max\! \,_{k=1}^{M} w_k.
  \label{eq:threshold}
\end{equation}
The threshold mimics the experimental reality that small peaks cannot
be distinguished from the noise, and, hence, have their $g$-factors
left out in the statistical analysis. Further, omitting
$g$-factors for which $w_{k} < w_{\rm tr}$
enforces the ``selection rules'' in the absence of
spin-orbit scattering. We verified that the precise
definition of $w_{\rm tr}$ does not affect our conclusions.

\begin{figure}[t]
\epsfxsize= 0.75\hsize
\epsffile{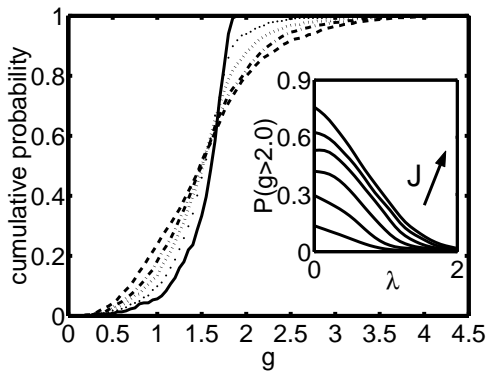}


\caption{\label{probability_distribution_exp}
\label{fig:prob_g>2}
Cumulative $g$-factor distribution
for $\lambda = 0.70$, $J=0$ (solid curve), 
$\lambda = 0.85$, $J=0.2\spacing$ (points), 
$\lambda=0.90$, $J=0.3\spacing$ (dotted), 
$\lambda=1.0$,d and $J=0.4\spacing$ (dash-dot),
and $\lambda=1.15$, $J=0.6\spacing$ (dashed).
The values of $\lambda$ are chosen such that
$\langle g \rangle=1.58$ in all cases. 
Inset: Probability for a level 
to have a $g$-factor larger than two, for $J/\spacing=0.1 - 0.6$, 
versus $\lambda$. }
\end{figure}

Interaction effects increase the ensemble averaged $g$-factor $\langle
g \rangle$ significantly for $J \gtrsim 0.3 \spacing$ and $\lambda
\lesssim 2$, see Fig.\ \ref{weighted_average_g_factors}.
In fact, there is a substantial parameter window for which $\langle g
\rangle > 2$.  By itself, such an increase of $\langle g \rangle$ has
limited experimental relevance, since there is no independent method
to measure $\lambda$. In fact, comparison of the measured
$\langle g \rangle$ with theory is used to determine the spin-orbit
rate $1/\tau_{\rm so} = \lambda^2 \spacing/\pi$ in nanoparticles
\cite{kn:petta2001,kn:petta2002}. This problem does not exist for the
full (cumulative) probability distribution of $g$-factors (average and
fluctuations), which is shown in
Fig.~\ref{probability_distribution_exp}. The values of $J$ and 
$\lambda$ in Fig.\
\ref{probability_distribution_exp} are chosen such that all
distributions have the same average $\langle g \rangle$.
The distributions of Fig.\ \ref{probability_distribution_exp},
together with our numerical results for other values of $J$ and
$\lambda$ (not shown), show that the exchange interactions 
substantially enhance the $g$-factor fluctuations (at the same
value of the average). 
The probability $P(g>2)$ to find a $g$-factor larger than two
is shown in the inset of Fig.\ \ref{fig:prob_g>2}, as a function of
$J$ and $\lambda$.

Presently, $g$-factor distributions have been measured for the 
noble metals only \cite{kn:petta2001,kn:petta2002}, for 
which $J/\delta \lesssim 0.1$ and interaction effects are 
negligible \cite{kn:macdonald1982}. 
Indeed, the distributions measured in Refs.\
\onlinecite{kn:petta2001,kn:petta2002} are in good agreement
with the non-interacting theory \cite{kn:brouwer2000,kn:matveev2000}.
Interaction effects, nontrivial spin states, and, hence,
$g$-factors larger than two should be observable for most other metals.
For alkali metals, $J/\spacing$ is in the range $0.2$--$0.3$
\cite{kn:knecht1975}, as well for Ti, Zr, and Mg 
\cite{kn:bakonyi1993,kn:wilk1978}. Even stronger interaction
effects are expected for Nb, Rh, and Y nanoparticles, for 
which spin-density calculations set $J/\spacing$ around $0.4$
\cite{kn:sigalas1994}, and for Pt, Pd, and V, which have 
$0.6 \lesssim J/\spacing < 1$ \cite{kn:fradin1975,kn:stenzel1986}.
For particles in the nm size range, spin-orbit effects are expected 
to be moderate or weak (dimensionless spin-orbit
rate $\lambda$ of order unity or smaller), except for the elements
with the highest atomic numbers (Au, Pt). (Measured spin-orbit rates in
Ag and Cu nanoparticles with radius $\sim 4$ nm
were in the range $\lambda \sim 1$
\cite{kn:petta2001}.) For those elements for which spin-orbit effects
are too weak to make large $g$-factors visible, spin-orbit scattering
can be enhanced by doping with a small amount of, {\em e.g.,} Au 
atoms \cite{kn:salinas1999}. Similarly, doping with ferromagnetic
atoms may increase $J$ and drive the metals towards
the Stoner instability at $J/\spacing = 1$ \cite{foot3}. 

In conclusion, we have shown that the presence of an exchange
interaction with strength
$J \gtrsim 0.2 \spacing$ leads to a significant broadening 
of the probability distribution of tunneling spectroscopy
$g$-factors in normal metal
nanoparticles. In particular, $g$-factors larger than two can be
observed, which are a signature of nontrivial many-electron
states. Weak spin-orbit scattering ($1/\tau_{\rm so} 
\lesssim \spacing$) is crucial in rendering the large
$g$-factors observable, since it mixes many-electron states with
different spin and, hence, lifts spin selections rules. 
It is only at larger spin-orbit scattering rates $1/\tau_{\rm so} 
\gg \spacing$ that spin-orbit scattering fully randomizes the spin 
and suppresses Zeeman contribution to $g$-factors and the 
matrix elements of the exchange interaction. The interaction
range $J/\spacing \gtrsim 0.2$ is appropriate for most metals, 
except for Al and the noble metals, which have negligible
exchange interaction effects.

Whereas in gated semiconductor
quantum dots the existence of nontrivial
spin states could be inferred indirectly
from the statistical distribution of Coulomb-blockade peak spacings
\cite{kn:usaj2002}, a comparison of the detailed parameter-dependence
of successive peak positions \cite{kn:luescher2001}, or from
the succession pattern of $g$-factors of many consecutive levels
\cite{kn:folk2001}, no such methods are available in metal
nanoparticles without a gate or with a very limited gate voltage
range. We hope that our finding that spin-orbit scattering renders
nontrivial $g$-factors visible in standard
tunneling spectroscopy without a gate motivates
further experiments on metal nanoparticles with strong interaction
effects.

We thank S.~Adam, A.~Kaminski, A.~H.~MacDonald,
J.~R.~Petta and D.~C.~Ralph 
for helpful discussions.
This work was supported by the NSF under grant 
no.\ DMR 0086509 and by the Packard foundation.

\vspace{-0.1cm}


\begin{thebibliography}{28}
\bibitem[{foo({\natexlab{b}})}]{foot0}
\bibinfo{note}{The other contribution is a ``pairing interaction'', which, if
  attractive, causes a superconducting instability at sufficiently low
  temperatures. We consider normal metal nanoparticles for which the pairing
  interaction is small}.
\end{thebibliography}

\end{document}